\documentstyle[twocolumn]{article}
\begin{document}
\title{\Large\bf Sonoluminescence as quantum vacuum radiation}
\author{{\large Claudia Eberlein}\\
        {\normalsize\em Department of Physics, University of Illinois
	at Urbana-Champaign,  Urbana, Illinois 61801-3080, U.S.A.}\\
	{\normalsize (Submitted May 30, 1995)}}
\date{\parbox{140mm}{\small
\hspace*{3mm} Sonoluminescence is explained in terms of quantum radiation by
moving interfaces between media of different polarizability. It can be
considered as a dynamic Casimir effect, in the sense that it is a
consequence of the imbalance of the zero-point fluctuations of the
electromagnetic field during the non-inertial motion of a
boundary. The transition amplitude from the vacuum into a two-photon
state is calculated in a Hamiltonian formalism and turns out to be
governed by the transition matrix-element of the radiation pressure.
Expressions for the spectral density and the total radiated energy are
given.\\ 
\\
PACS numbers: 03.70.+k, 11.10.-z, 42.50.Lc, 78.60.Mq}}
\maketitle

Sonoluminescence is a phenomenon that has so far resisted all attempts
of explanation. A short and intense flash of light is observed when
ultrasound-driven air or other gas bubbles in water collapse. This
process has been known for more than 60 years to occur randomly when
degassed water is irradiated with ultrasound \cite{first}. Recently
interest has been revived by the contriving of stable sonoluminescence
\cite{stable1,stable2} where a bubble is trapped at the pressure
anti-node of a standing sound-wave in a spherical or cylindrical
container and collapses and re-expands with the periodicity of the
sound. With a clock-like precision a light pulse is emitted during
every cycle of the sound-wave; the jitter in the sequence of pulses is
almost unmeasurably small. Shining laser light upon the bubble and
analyzing the scattered light on the basis of the Mie theory of
scattering from spherical obstacles one has been able to record the
time-dependence of the bubble radius \cite{miemeas}; these experiments
showed that the flash of light is emitted shortly after the bubble has
collapsed, i.e. shortly after it has reached its minimum radius. This
and the fact that the spectrum of the emitted light resembles
radiation from a black body at several tens of thousands degree Kelvin
have led to the conjecture that the light could be thermal radiation
from the highly compressed and heated gas contents of the bubble after
the collapse \cite{spec}. It has also been argued that the
experimentally observed spectrum would equally well be compatible with
the idea of a plasma forming at the bubble centre after the collapse
and radiating by means of bremsstrahlung \cite{brems}. An alternative
suggestion has tried to explain the sonoluminescence spectrum as
pressure-broadened vibration-rotation lines \cite{susl}, but although
this theory has been very successful in the case of randomly excited
(multi-bubble) sonoluminescence seen in silicone oil it has been
inefficacious for sonoluminescence in water.

All of the above theories have serious flaws. Both black-body
radiation and bremsstrahlung would make a substantial part of the
radiated energy appear below 200nm where the surrounding water would
absorb it. If one estimates the total amount of energy to be absorbed
corresponding to the observed number of photons above the absorption
edge, one quickly convinces oneself that this would be far too much to
leave no macroscopically discernible traces in the water, as for
instance dissociation \cite{priv}; however, nothing the like is
observed. Another very strong argument against all three of the above
theories is that the processes involved in each of them are far too
slow to yield pulse lengths of 10ps or less but which are
observed. Moreover, if a plasma were formed in the bubble, one should
see at least remnants of slow recombination radiation from the plasma
when the bubble re-expands. As to the theory involving
vibration-rotation excitations, the line-broadening required to model
the observed spectrum seems rather unrealistic.

In its concept the theory to be presented here has been loosely
inspired by Schwinger's idea \cite{jschw} that sonoluminescence might
be akin to the Casimir effect in the sense that the zero-point
fluctuations of the electromagnetic field might lie at the origin of
the observed radiation. More closely related to this is the Unruh
effect well-known in field theory \cite{unruh}; its original statement
is that a uniformly accelerated mirror in vacuum emits photons with
the spectral distribution of black-body radiation. However, the
phenomenon is far more general than that and in particular not
restricted to perfect mirrors. This kind of quantum vacuum radiation
has been shown to be generated also by moving dielectrics
\cite{GBCE}. Whenever an interface between two dielectrics or a
dielectric and the vacuum moves non-inertially photons are created. In
practice this effect is very feeble, so that it has up to now been far
beyond any experimental verification. Sonoluminescence might be the
first identifiable manifestation of quantum vacuum radiation.

The mechanism by which radiation from moving dielectrics and mirrors
in vacuum is created is understood most easily by picturing the medium
as an assembly of dipoles. The zero-point fluctuations of the
electromagnetic field induce these dipoles and orient and
excite them. However, as long as the dielectric stays stationary or
uniformly moving such excitations remain virtual; real photons are
created only when the dielectric or mirror moves non-uniformly,
because then the fluctuations get out of balance and no longer average
to zero. Mechanical energy of the motion of the dielectric is
dissipated into the field and a corresponding frictional force is felt
by the dielectric. The fluctuation-dissipation theorem predicts this
frictional force \cite{CEfluc} in terms of the force fluctuations on
the stationary dielectric or mirror \cite{BEfluc}; it holds, however,
no information on the state of the photon field, i.e. the radiated
spectrum cannot be evaluated from the fluctuation-dissipation theorem.

The surface of an air bubble in water is such an interface between two
dielectric media. When the bubble collapses, the motion of the
interface is highly non-linear; the acceleration and higher
derivatives of the velocity attain values that are high enough to make
quantum vacuum radiation a non-negligible process.

The present model describes the bubble as a spherical cavity in a
uniform dielectric medium. The refractive index of water is roughly
1.3 in the visible spectrum, and the air inside the bubble has a
refractive index of practically 1 even if strongly compressed. The
assumption of the uniformity of the water is of course unrealistic,
but the variation of the refractive index in the vicinity of the
bubble surface is of secondary importance for the problem of vacuum
radiation. For the present purposes the bubble can to a very good
approximation be described by a step in the dielectric function
\begin{equation}
        \varepsilon(r;R) = 1 + (n^2 -1)\: \theta(r-R)\;.
\label{eps}
\end{equation}
Here $n$ is the refractive index of water which is for simplicity
assumed to be constant and non-dispersive, and the refractive index of
the bubble contents has been set to 1.

The step in $\varepsilon$ imposes continuity conditions on the
components of the electric displacement vector $\bf D$ and the
magnetic field $\bf B$ at the bubble surface; this provides the
coupling between the fields and the motion of the bubble. The latter
is described by the time-dependence of the bubble radius $R(t)$ which
is in the present model taken to be an externally prescribed
parameter; the hydrodynamics of the bubble motion is not the concern
of this letter. However, an expression for the frictional force that
is due to the momentum transfer from the mechanical degrees of freedom
of the bubble motion into the field is obtained as one of the end
results and ought to be taken into consideration in the hydrodynamic
equations of motion of the bubble \cite{hydro}.

The dynamics of the electromagnetic fields is classically as well as
quantally described by the Hamiltonian
\begin{equation}
        H = \int{\rm d}^3{\bf r} \left[ \frac{1}{2}\left(
        \frac{{\bf D}^2}{\varepsilon} + {\bf B}^2 \right) +
        \beta \frac{\varepsilon -1}{\varepsilon}\;
        ({\bf D}\wedge{\bf B})_r \right]\,,
\label{ham}
\end{equation}
where $\beta$ is the velocity of the bubble surface in units of the
speed of light; the subscript $r$ denotes the radial component of a
vector with respect to the centre of the bubble. The first part of $H$
is the usual Hamiltonian for a stationary dielectric; the second part
is a motional correction which is small by virtue of $\beta$ being
small. This Hamiltonian has been derived from considerations of
Lorentz invariance \cite{tobe}; acceleration stresses have been
neglected, and so have terms of order $\beta^2$ and higher.

The transition amplitude for the photon field to go from its initial
vacuum to an excited state is calculated by solving the
Schr\"{o}dinger equation
\begin{equation}
        {\rm i}\:\hbar\,\frac{\rm d}{{\rm d}t}\,|\psi\rangle =
        H |\psi\rangle\;
\end{equation}
with the initial condition $|\psi(t_0)\rangle = |0\rangle$. A
perturbative solution of this equation to first order in the interface
velocity $\beta$ is called for. This poses a non-trivial problem,
since the Hamiltonian $H$ depends not only explicitly on
$\dot{R}(t)\equiv\beta$ but via $\varepsilon$ also parametrically on
$R(t)$. To handle this task a judicious combination of standard
perturbation theory and Pauli's theory of adiabatic approximation has
been devised \cite{thesis,tobe}; its application yields for the
transition amplitude from the initial vacuum $|0\rangle$ into a
two-photon state $|k,k'\rangle$ to first order in $\beta$
\begin{eqnarray}
   \langle k,k'|\psi\rangle&\! =&\! - \frac{1}{\omega +\omega'}\:
	\int_{t_0}^{t} {\rm d}\tau\;\beta(\tau)\:
	{\rm e}^{{\rm i}(\omega+\omega')(\tau-t)}
	\nonumber\\
        &&\hspace*{20mm}
	\times\langle k,k'|\;{\cal F}_r\,|0\rangle_{R(\tau)}\;.
\label{ampl}
\end{eqnarray}
where the matrix element of ${\cal F}_{r}$ has to be taken at the
bubble radius $R(\tau)$. The operator ${\cal F}_{r}$ is defined by
\begin{eqnarray}
        {\cal F}_{r}&\! =&\! - \left( 1-\frac{1}{n^2} \right)
        \frac{R^2}{2} \oint {\rm d}\Omega \nonumber\\
	&& \times \left[ \left(
        1+\frac{1}{n^2}\right) D_r^2 + B_r^2 - B_\theta^2
        - B_\phi^2 \right]\,.
\label{force}
\end{eqnarray}
One of the most intriguing results of this calculation is that ${\cal
F}_{r}$ is not merely a shorthand for an integral over squared field
components, but turns out to have a physical meaning; it is the radial
component of the force exerted by the field onto the interface. This
shows that the emission of photons by a moving dielectric is indeed
intrinsically related to the zero-point fluctuations of the radiation
pressure. This relation can be made even more transparent by
considering the mean-square deviation of the force on the surface of a
stationary bubble.
\begin{eqnarray}
\triangle F_r^{2}&\! =&\! \langle 0 | {\cal F}_r^{2} | 0 \rangle
                    - \langle 0 | {\cal F}_r | 0 \rangle^{2}
\nonumber\\
		 &\! =&\! \frac{1}{2}\: \int{\rm d}k \int{\rm d}k' \;
                    | \langle 0 | {\cal F}_r | k, k' \rangle |^{2} \;.
\label{deltaF}
\end{eqnarray}
The last expression is derived by inserting an identity operator
decomposed into the complete set of photon eigenstates; as ${\cal
F}_{r}$ is an operator that is quadratic in the fields only two-photon
states give non-zero matrix elements. These virtual two-photon states
become real when the system is perturbed, which in this case means when
the dielectric starts moving. Eq. (\ref{ampl}) reveals that the
spectrum of the emitted photons is determined by the spectrum of the
zero-point fluctuations of the field. As discussed above the
fluctuation-dissipation theorem underlies this fundamental
interrelation, although it does not exhaust it.

In principle the transition amplitude (\ref{ampl}) allows one to
calculate all physically significant quantities concerning the
radiation process. Experimentally most important is the angle-integrated
spectral density
\begin{equation}
        {\cal P}(\omega) = \omega^3\int_0^T {\rm d}t
        \oint {\rm d} \Omega_{\bf k}\;
        \int_{-\infty}^{\infty}{\rm d}^3 {\bf k}'\;
        |\langle k,k'|\psi\rangle|^{2}\;.
\label{spectr}
\end{equation}
${\cal P}(\omega)$ is a functional of the trajectory $R(t)$ of the
bubble surface. Its direct analytical determination is hindered by the
multiple occurrence of $R(\tau)$ in $\langle k,k'|\;{\cal F}_r\,|0
\rangle_{R(\tau)}$ and by the complicated structure of this matrix
element which comprises products of spherical Bessel functions and
their derivatives \cite{foot}. To estimate the spectral density
$P(\omega)$ one can adopt a model profile for the time-dependence of
the bubble radius about the collapse
\begin{equation}
        R^2(t) = R_0^2 - \left(R_0^2 - R_{\rm min}^2\right) \,
        \frac{1}{(t/\gamma)^2 + 1}\;.
\label{modR}
\end{equation}
$R_0$ and $R_{\rm min}$ are the initial and minimum radii,
respectively; the parameter $\gamma$ describes how fast the collapse
happens or, in other words, characterizes the time-scale of the
turn-around of the velocity $\beta(t)\equiv\dot{R}(t)$ at $R_{\rm
min}$. Assuming for feasibility furthermore that the bubble radius $R$
is much greater than the wavelengths of the light emitted \cite{foot},
one can derive for the spectral density
\begin{equation}
        {\cal P}(\omega) = \frac{(n^2-1)^2}{64\,n^2} \,
        \frac{\hbar}{c^4\gamma}\,
        \left( R_0^2 - R_{\rm min}^2\right)^2\:
        \omega^3\:{\rm e}^{-2\gamma\omega}\;.
\label{specdens}
\end{equation}
This is a result of great significance as it shows that the spectrum
of the emitted light resembles a black-body spectrum although
zero-temperature quantum field theory is being dealt with. The reason
for that lies in the nature of the zero-point fluctuations of the
electromagnetic field. Since the Hamiltonian is quadratic in the
fields, the photons are always created in pairs. The spectral density,
however, is determined in a single-photon measurement which involves
the tracing over the other photon in the pair; such tracing is known
to make pure two-mode states look like thermally distributed
single-mode states \cite{knight}.

Another quantity of interest is the total energy $\cal W$ radiated
during one acoustic cycle. In the short-wavelengths limit \cite{foot}
one obtains
\begin{equation}
        {\cal W} = \frac{(n^2-1)^2}{n^2}\,\frac{\hbar}{480\pi c^3}
        \int_{0}^{T} {\rm d}\tau\; \frac{\partial^5 R^2(\tau)}{\partial
        \tau^5}\,R(\tau)\beta(\tau)\;.
\label{Wres}
\end{equation}
{}From this the dissipative force acting on the moving bubble surface is
seen to behave like $R^2\beta^{(4)}$ in its leading term. Such a
proportionality to the fourth derivative of the velocity is also found
in calculations of frictional forces on moving plane perfect mirrors
\cite{paulo}. The emission of photons is thus not predominantly
influenced by the acceleration of the interface, which retrospectively
justifies the disregard of acceleration stresses in the present model.

A reckoning based on the model trajectory (\ref{modR}) yields
the estimate
\begin{equation}
        {\cal W} = 2\cdot 10^{-13}\,{\rm J}\ \ {\rm for}\
        \gamma\sim 1{\rm fs}\,,
\label{numW}
\end{equation}
which corresponds roughly to the observed number of photons. One
femtosecond seems a very short time-scale for the turn-around of the
velocity, but numerical calculations \cite{tobe} indicate that the
photon emission is substantially enhanced by resonances in the regime
$kR\sim 1$, i.e.  when the photon wavelengths are comparable to the
bubble radius, so that in practice a turn-around time of the order of
100fs is presumably sufficient to yield the above amount of energy per
burst.

In conclusion, it can be said that at this crude level of inspection
the theory of vacuum radiation seems to agree remarkably well with the
experimental results on sonoluminescence. The amount of the radiation
and the thermal-like spectrum are returned by the theory and further
numerical investigations will uncover more details. Likewise one has
no difficulties explaining the shortness of the observed pulses. The
pulse length is dictated by the time it takes for the zero-point
fluctuations to correlate around the bubble and by the turn-around
time of the velocity about the collapse; both are much shorter than
10ps. Another major point that is clarified by this theory is that
there are practically no photons created in the UV and at even higher
energies, as water has no appreciable polarizability there. Hence no
radiation has to be absorbed by the surrounding water.

A relatively simple experiment to discriminate the present from other
theories of sonoluminescence is to look for photons radiated in the
X-ray transparency window of water \cite{priv}; whereas both
black-body and bremsstrahlung theories predict a perceptible number of
photons radiated into this window, the theory of vacuum radiation
forbids them.

A second, not too difficult distinguishing experiment is to measure
the angular distribution of the light emitted from an elongated rather
than spherical sonoluminescent bubble achieved by anisotropic
ultrasound. The present theory, unlike others, predicts an anisotropic
sonoluminescence intensity, as the number of photons emitted into a
certain direction is roughly proportional to the cross-section of the
bubble perpendicular to this direction.

Vacuum radiation might strike one as a strange explanation for the
light seen in sonoluminescence, since one often tends to think of
low-energy photons as emitted by atoms. However, the present case
forces one to give up this lax point of view, as atomic transitions
are about a thousand times slower than a sonoluminescence pulse. On
the level of quantum electrodynamics radiation comes from moving
charges and it lies within one's discretion whether one groups these
charges in atoms, in dipoles to make up a dielectric, or in yet
another structure.  For sonoluminescence it seems most convenient to
think in terms of a dielectric as a whole in order to account for the
cooperative response of charges to the zero-point fluctuations of the
electromagnetic field.

It is a pleasure to thank Peter W. Milonni for drawing my attention to
sonoluminescence. Stimulating discussions with Gabriel Barton, Paul
M. Goldbart, Nigel D. Goldenfeld, Peter L. Knight, Anthony J. Leggett,
Efrat Shimshoni, and Shivaji L. Sondhi are gratefully acknowledged.
David K. Campbell is thanked for his encouraging interest in this work.
This research has been supported through the John D. and Catherine
T. MacArthur Foundation.


\begin{thebibliography}{99}
\bibitem{foot}{Calculations concerning photon scattering in spherical
coordinates involve spherical Bessel functions $j_\ell(kR)$ and
$y_\ell(kR)$. In the short-wavelengths limit, i.e. when $kR\gg1$,
dominant contributions come from large angular momenta $\ell$ but
$\ell\leq kR$, so that uniform asymptotic expansions can be employed.
If, however, $kR\sim 1$ these approximations are rendered invalid and
estimations can be conducted only numerically.}
\end{thebibliography}
\end{document}